\documentclass[showpacs,twocolumn,aps]{revtex4}
\usepackage{amssymb}
\usepackage{amsmath}
\usepackage{graphicx}
\usepackage{lscape}
\begin{document}
\title{Centrality bin size dependence of multiplicity correlation \\in central Au+Au
       collisions at $\sqrt{s_{\rm{NN}}}$=200 GeV}
\author{Yu-Liang Yan$^{1}$, Dai-Mei Zhou$^{2}$, Bao-Guo Dong$^{1,3}$, Xiao-Mei Li$^{1}$,
Hai-Liang Ma$^{1}$, Ben-Hao Sa$^{1,2,4}$\footnote{Corresponding author:
sabh@ciae.ac.cn}}
\address{1 China Institute of Atomic Energy, P.O. Box 275(18), Beijing 102413, China \\
2 Institute of Particle Physics, Huazhong Normal University, Wuhan 430079, China\\
3 Center of Theoretical Nuclear Physics, National Laboratory of Heavy Ion Collisions,
Lanzhou 730000,China\\
4 CCAST (World Laboratory), P. O. Box 8730 Beijing 100080, China}

\begin{abstract}
We have studied the centrality bin size dependence of charged particle
forward-backward multiplicity correlation strength in 5\%, 0-5\%, and 0-10\% most
central Au+Au collisions at $\sqrt{s_{\rm{NN}}}$=200 GeV with a parton and hadron
cascade model, PACIAE based on PYTHIA. The real (total), statistical, and NBD
(Negative Binomial Distribution) correlation strengths are calculated by the real
events, the mixed events, and fitting the charged particle multiplicity distribution
to the NBD, respectively. It is turned out that the correlation strength increases
with increasing centrality bin size monotonously. If the discrepancy between real
(total) and statistical correlation strengths is identified as dynamical one, the
dynamical correlation may just be a few percent of the total (real) correlation.
\end{abstract}

\maketitle

\section{INTRODUCTION}
The study of fluctuations and correlations has been suggested as a useful means for
revealing the mechanism of particle production and Quark-Gluon-Plasma (QGP) formation
in Relativistic Heavy Ion Collisions \cite{hwa2,naya}. Correlations and fluctuations of
the thermodynamic quantities and/or the produced particle distributions may be
significantly altered when the system undergoes phase transition from hadronic matter to
quark-gluon matter because the degrees of freedom in two matters is very different.

The experimental study of fluctuations and correlations becomes a hot topic in
relativistic heavy ion collisions with the availability of high multiplicity
event-by-event measurements at the CERN-SPS and BNL-RHIC experiments. An abundant
experimental data have been reported \cite{star2,phen2,phob} where a lot of new
physics arise and are urgent to be studied. A lot of theoretical
investigations have been reported as well \cite{paja,hwa1,konc,brog,fu,yan,bzda}.

Recently STAR collaboration have measured the charged particle forward-backward
multiplicity correlation strength $b$ in Au+Au collisions at $\sqrt{s_{\rm{NN}}}$=200
GeV \cite{star3,star4}. The outstanding features of STAR data are:
\begin{itemize}
\item In most central collisions, the correlation strength $b$ is approximately flat
across a wide range in $\Delta \eta$ which is the distance between the centers of
forward and backward (pseudo)rapidity bins.
\item This trend disappears slowly with decreasing centrality and approaches a
exponential function of $\Delta \eta$ at the peripheral collisions.
\end{itemize}
That has stimulated a lot of theoretical interests \cite{konc,brog,fu,bzda}.

In Ref. \cite{fu}, a statistical model was proposed to calculate the charged particle
forward-backward multiplicity correlation strength $b$ in 0-10\% most central Au+Au
collisions at $\sqrt{s_{\rm{NN}}}$=200 GeV. One outstanding feature of STAR data,
the $b$ as a function of $\Delta \eta$ is approximately flat, was well reproduced. The
calculated value of $b\approx0.44$ was compared with STAR data of $b\approx0.60$
\cite{star3,star4}.

However, in this statistical model \cite{fu} the Negative Binomial Distribution (NBD)
is assumed for the charged multiplicity distribution and the NBD parameters of $\mu$
and $k$ (see later) are extracted from fit in with PHENIX charged particle
multiplicity distribution \cite{phen3}. It is turned out in Ref. \cite{yan}
that the experimental $\eta$ and $p_T$ acceptances have large influences on the
correlation strength $b$. The STAR experimental acceptances are quite different from
PHENIX, thus the inconsistency, using PHENIX multiplicity data to explain STAR
correlation data, involved in \cite{fu} have to be studies further. Meanwhile, what is
the discrepancy between $b\approx0.60$ (STAR datum) and $b\approx0.44$ (NBD) also
needs to be answered.

In this paper we use a parton and hadron cascade model PACIAE \cite{sa}, to
investigate the centrality bin size dependence of charged particle multiplicity
correlation in 5, 0-5, and 0-10\% most central Au+Au collisions at $\sqrt{s_{ \rm{NN}
}}$=200 GeV. Following Ref. \cite{yan} we generate the real events (6000) by the
PACIAE model, construct the mixed events according to real events one by one, and
extract the NBD parameters ($\mu$ and $k$) from fitting the real events charged
particle multiplicity distribution to the NBD. Then the charged particle
forward-backward multiplicity correlation strength $b$ is calculated for the real
events (real correlation strength), the mixed events (statistical correlation
strength), and the NBD (NBD correlation strength), respectively. They are all nearly
flat across a widw range in $\Delta \eta$. Their magnitude in 0-10\% most central Au+Au
collisions are about 0.63, 0.59, and 0.52, respectively. So the corresponding STAR data
is well reproduced. It is turned out that the real (total), statistical, and NBD
correlation strengths increase with increasing centrality bin size monotonously.
If the discrepancy between real (total) and statistical correlation strengths is
identified as dynamical one, then the dynamical correlation strength may just be a few
percent of the total (real) correlation strength.

\section {PACIAE MODEL}
The parton and hadron cascade model, PACIAE \cite{sa}, is based on PYTHIA \cite{soj2}
which is a model for hadron-hadron ($hh$) collisions. The PACIAE model is composed of
four stages: parton initialization, parton evolution (rescattering), hadronization,
and hadron evolution (rescattering).

\subsection {Parton initialization}
In the PACIAE model a nucleon-nucleon ($NN$) collision is described with PYTHIA model,
where a $NN$ ($hh$) collision is decomposed into the parton-parton collisions. The
hard parton-parton collision is described by the lowest-leading-order (LO) pQCD
parton-parton cross section \cite{comb} with modification of parton distribution
function in the nucleon. And the soft parton-parton interaction is considered
empirically. The semihard, between hard and soft, QCD $2\rightarrow 2$ processes are
also involved in PYTHIA (PACIAE) model. Because of the initial- and final-state QCD
radiation added to the above processes, the PYTHIA (PACIAE) model generates a
multijet event for a $NN$ ($hh$) collision. That is followed, in the PYTHIA model, by
the string-based fragmentation scheme (Lund model and/or Independent Fragmentation
model), thus a hadronic state is reached for a $NN$ ($hh$) collision. However, in the
PACIAE model above string fragmentation is switched off temporarily, so the result
is a multijet event (composed of quark pairs, diquark pairs and gluons) instead of a
hadronic state. If the diquarks (anti-diquarks) are split forcibly into quarks
(anti-quarks) randomly, the consequence of a $NN$ ($hh$) collision is its initial
partonic state composed of quarks, anti-quarks, and gluons.

A nucleus-nucleus collision, in the PACIAE model, is decomposed into the
nucleon-nucleon collisions based on the collision geometry. A nucleon in the colliding
nucleus is randomly distributed in the spatial coordinate space according to the
Woods-Saxon distribution ($r$) and the 4$\pi$ uniform distribution ($\theta$ and
$\phi$). The beam momentum is given to $p_z$ and $p_x=p_y=0$ is assumed for each
nucleon in the colliding nucleus. A closest approaching distance of two assumed
straight line trajectories is calculated for each $NN$ pair. If this distance is less
than or equal to $\displaystyle{\sqrt{\sigma_{\rm{tot}}/\pi}}$, then it is considered
as a collision pair. Here $\sigma_{\rm{tot}}$ refers to the total cross section of
$NN$ collision assumed to be 40 mb. The corresponding collision time of this collision
pair is then calculated. So the particle list and the $NN$ collision (time) list can
be constructed. A $NN$ collision pair with smallest collision time is selected from
the $NN$ collision (time ) list and performed by the method in former paragraph. After
upgrading the particle list and collision (time) list we select and perform a new $NN$
collision pair again. Repeat these processes until the collision (time) list is empty
we obtain a initial partonic state for a nucleus-nucleus collision.

\subsection {Parton evolution (rescattering)}
The next step, in the PACIAE model, is parton evolution (partonic rescattering).
Here the $2\rightarrow2$ LO-pQCD differential cross sections \cite{comb} are employed.
The differential cross section for a subprocess $ij\rightarrow kl$ reads
\begin{equation}
\frac{d\sigma_{ij\rightarrow
kl}}{d\hat{t}}=K\frac{\pi\alpha_s^2}{\hat{s}}\sum_{ij\rightarrow
kl},
\end{equation}
where the $K$ factor is introduced for higher order corrections and the
non-perturbative QCD correction as usual. Take the process $q_1q_2 \rightarrow q_1q_2$
as an example, one has
\begin{equation}
\sum_{q_1q_2\rightarrow
q_1q_2}=\frac{4}{9}\frac{\hat{s}^2+\hat{u}^2}{\hat{t}^2},
\label{eq3}
\end{equation}
which can be regularized as
\begin{equation}
\sum_{q_1q_2\rightarrow
q_1q_2}=\frac{4}{9}\frac{\hat{s}^2+\hat{u}^2}{(\hat{t}-m^2)^2}
\end{equation}
by introducing the parton colour screen mass, $m$=0.63 GeV. In above equation
$\hat{s}$, $\hat{t}$, and $\hat{u}$ are the Mandelstam variables and $\alpha_s$=
0.47 stands for the running coupling constant. The total cross section of the parton
collision $i+j$ is
\begin{equation}
\sigma_{ij}(\hat{s})=\sum_{k,l}\int_{-\hat{s}}^{0}d\hat{t}
\frac{d\sigma_{ij\to kl}}{d\hat{t}}.
\end{equation}
With these total and differential cross sections the parton evolution (rescattering)
can be simulated by the Monte Carlo method until the parton-parton collision is
ceased (partonic freeze-out).

\subsection {Hadronization}
In the PACIAE model the partons can be hadronized with the string-based fragmentation
scheme or by the coalescence (recombination) models \cite{biro,csiz,grec,hwo,frie}. The
Lund string fragmentation regime, involved in the PYTHIA model, is adopted for
hadronization in this paper, see \cite{soj2} for the details.

Meanwhile, we have proposed a simulant coalescence (recombination) model which can be
briefly explained as follows:
  \begin{enumerate}
  \item The Field-Feynman parton generation mechanism \cite{ff1} is first applied
to deexcite the energetic parton and thus to increase the parton multiplicity. This
deexcitation of an energetic parton plays a similar role as string multiple
fragmentation in the Lund model \cite{and1}.
  \item The gluons are forcibly split into $q\bar q$ pair randomly.
  \item In the program there is a hadron table composed of mesons and baryons. The
pseudoscalar and vector mesons made of u, d, s, and c quarks, as well as $B^+$,
$B^0$, $B^{*0}$, and $\Upsilon$ are considered. The SU(4) multiplets of
baryons made of u, d, s, and c quarks (except those with double c quarks) as well
as $\Lambda^0_b$ are considered.
  \item Two partons can coalesce into a meson and three partons into a baryon
(antibaryon) according to the flavor, momentum, and spatial coordinates of partons and
the valence quark structure of hadron.
  \item When the coalescing partons can form either a pseudoscalar meson or a vector
meson (e. g. $u\bar d$ can form either a $\pi^+$ or a $\rho^+$) a judgment of less
discrepancy between the invariant mass of coalescing partons and the mass of coalesced
hadron is invoked to select one from two mesons above. In the case of baryon, e. g. both
$p$ and $\Delta^+$ are composed of $uud$, the same judgment is invoked to select one
baryon from both of $\frac{1}{2}^+$ and $\frac{3}{2}^+$ baryons.
  \item Four momentum conservation is required.
  \item There is a phase space condition
  \begin{equation}
  \frac{16\pi^2}{9}\Delta r^3\Delta p^3=\frac{h^3}{d},
  \end{equation}
where $h^3/d$ is the volume occupied by a single hadron in the phase space, $d$=4
refers to the spin and parity degeneracies, $\Delta r$ and $\Delta p$ stand for the
spatial and momentum distances between coalescing partons, respectively.
  \end{enumerate}

\subsection {Hadron evolution (rescattering)}
We obtain a configuration of hadrons in spatial and momentum coordinate spaces for a
nucleus-nucleus collision after the hadronization. If one only considers the
rescattering among $\pi, k, p, n, \rho (\omega), \Delta, \Lambda, \Sigma, \Xi, \Omega,
J/\Psi$ and their antiparticles, the particle list is then constructed by the above
hadrons. A closest approaching distance of two assumed straight line trajectories is
calculated for each $hh$ pair. If this distance is less than or equal to
$\displaystyle{\sqrt{\sigma_{\rm{tot}}^{hh}/\pi}}$ \cite{sa1}, then it is considered
as a collision pair. Here $\sigma_{\rm{tot}}^{hh}$ refers to the total cross section
of $hh$ collision. The corresponding collision time of this collision pair is then
calculated. So the $hh$ collision (time) list can be constructed. A $hh$ collision
pair with smallest collision time is selected from the collision (time) list and
performed by the usual two-body collision method \cite{sa1}. After upgrading the particle
list and collision (time) list we select and perform a new $hh$ collision pair
again. Repeat these processes until the collision (time) list is empty (hadronic
freeze-out).

A isospin averaged parametrization formula is used for the $hh$ cross section
\cite{koch,bald}. However, we also provide a option of constant total, elastic, and
inelastic cross sections~\cite{sa1}: $\sigma_{\rm{tot}}^{NN}=40$~mb, $\sigma_{\rm{tot}
}^{\pi N}=25$~mb, $\sigma_{\rm{tot}}^{kN}=35$~mb, $\sigma_{\rm{tot}}^{\pi \pi}=10$~mb,
and the assumed ratio of inelastic to total cross section equals 0.85. We also assume
\begin{equation}
\sigma_{pp}=\sigma_{pn}=\sigma_{nn}=\sigma{\Delta N}=\sigma{\Delta \Delta}.
\end{equation}
The cross section of $\pi \bar N$ and $k\bar N$, for instance, is assumed to be equal to
the cross section of $\pi N$ and $kN$, respectively.

The momentum of scattered particles in a $hh$ elastic collision is simulated according
to that the $hh$ differential cross section, $d\sigma_{\rm{tot}}^{hh}/dt$, is assumed
to be an exponential function of $t$ which is squared momentum transfer \cite{sa1}. As
it is impossible to include all inelastic channels, we consider only a part of them
($\approx$ 600) which have noticeable effects on the hadronic final state, and the rest
is attributed to the elastic scattering. Take incident channel $\pi N$ as an example,
if there are possible final channels of $\pi N \rightarrow \pi \Delta $, $\pi N
\rightarrow \rho N$, and $\pi N\rightarrow k\Lambda$, their relative probabilities are
then used to select one among above three channels. The momentum of scattered particles
in a $hh$ inelastic collision is simulated according to the usual two-body kinematics
\cite{sa1,pdg}.

\section {CALCULATION AND RESULT}
Following \cite{cape} the charged particle forward-backward multiplicity correlation
strength $b$ is defined as
\begin{equation}
\ b =\frac{\langle n_fn_b\rangle - \langle n_f\rangle \langle
n_b\rangle}{\langle n_f^2\rangle - \langle n_f\rangle^2} =
\frac{{\rm {cov}}(n_f,n_b)}{{\rm{var}}(n_f)},
\end{equation}
where $n_f$ and $n_b$ are, respectively, the number of charged particles in forward and
backward pseudorapidity bins defined relatively and symmetrically to a given
pseudorapidity $\eta$. $\langle n_f\rangle$ refers to the mean value of $n_f$ for
instance. cov($n_f$,$n_b$) and var($n_f$) are the forward-backward multiplicity
covariance and forward multiplicity variance, respectively.

\begin{table}[htbp]
\caption{Total charged particle multiplicity in three $\eta$ fiducial ranges in 0-6\%
     most central Au+Au collisions at $\sqrt{s_{\rm{NN}}}$=200 GeV.}
\begin{tabular}{cccc}
\hline\hline
         & $N_{\rm{ch}}(|\eta|< 4.7)$  & $N_{\rm{ch}}(|\eta|< 5.4) $ & $N_{\rm{ch}}$(total)    \\
\hline
  PHOBOS$^a$ & 4810 $\pm$ 240  & 4960 $\pm$ 250 &  5060 $\pm$ 250                 \\
  PACIAE     & 4819  & 4983    &  5100    \\

\hline\hline
\multicolumn{4}{l}{$^a$ The experimental data are taken from \cite{phob2}.}\\
\end{tabular}
\label{mul}
\end{table}

In the calculations the default values given in the PYTHIA model are adopted for all
model parameters except the parameters $K$ and $b_s$ (in the Lund string fragmentation
function). The $K$=3 is assumed and the $b_s$=6 is fixed by fitting the charged
particle multiplicity to the corresponding PHOBOS data in 0-6\% most central Au+Au
collisions at $\sqrt{s_{\rm{NN}}}$=200 GeV \cite{phob2} as shown in Tab.~\ref{mul}.
Therefore in the generation of real events there is no free parameters. The mapping
relation \cite{sa2} between the centrality definition in theory and experiment
\begin{equation}
b_i=\sqrt{g}b_i^{\rm {max}},\qquad b_i^{\rm {max}}=R_A + R_B,
\label{imp}
\end{equation}
is employed. In the above equation $b_i$ (in fm) refers to the theoretical impact
parameter and $g$ stands for the percentage of geometrical (total) cross section
used in experiment to define the centrality. $R_A=1.12A^{1/3}+0.45$ fm, for
instance, is the radius of nucleus $A$. Thus the 0-10, 0-6, 0-5, and 5\% most central
collisions, for instance, are mapped to $0<b_i<4.46$, $0<b_i<3.53$, $0<b_i<3.20$, and
$b_i=3.20$ fm, respectively.

In this paper we propose a mixed event method where the mixed events are generated
according to real events one by one. We first assume the charged particle multiplicity
$n$ in a mixed event is the same as one corresponding real event. However,
$n$ particles of this mixed event are sampled randomly from the particle reservoir
composed of all particles in all real events. Therefore, there is no dynamical
relevance among the particles in a mixed event. So the correlation calculated by
mixed events is reasonably to be identified as the statistical correlation \cite{yan}.

It is known that the statistical correlation can also be studied by the NBD method,
because the charged particle multiplicity distribution in high energy heavy-ion
collisions is close to NBD \cite{phen3}. For an integer $n$ the NBD reads
\begin{equation}
\ P(n;\mu,k) = \begin{pmatrix}n+k-1\\k-1\end{pmatrix}\frac{
(\mu/k)^n}{(1+\mu/k)^{n+k}},
\label{NBD}
\end{equation}
where $\mu\equiv \langle n \rangle$ is a parameter, $P(n;\mu,k)$ is normalized in
$0\leq n \leq\infty$, and $k$ is another parameter responsible for the shape of the
distribution. As proved in \cite{yan} the correlation strength can be expressed as
\begin{equation}
\ b = \frac{\langle n_f\rangle}{\langle n_f\rangle+k},
\label{b3}
\end{equation}
where the parameter $k$ is fixed by fitting the charged particle multiplicity to the
NBD usually.

\begin{figure}[htbp]
\includegraphics[width=7cm,angle=0]{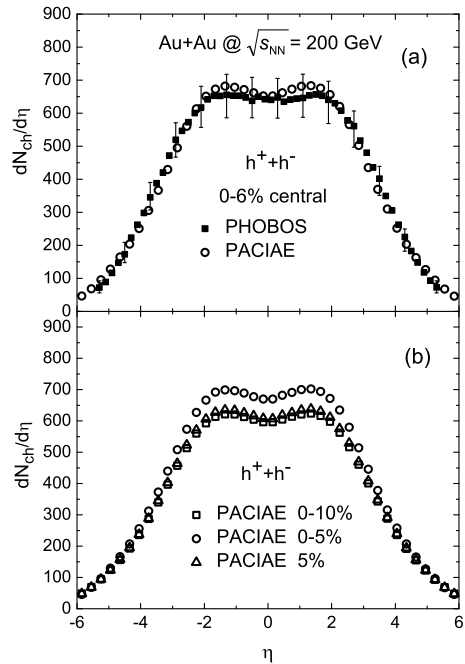}
\caption{Charged particle pseudorapidity distribution in Au+Au
collision at $\sqrt{s_{\rm{NN}}}$=200 GeV: (a) 0-6\% most central
collision and (b) 0-10, 0-5, and 5\% most central collision. The
experimental data are taken from \cite{phob2}.} \label{fig1}
\end{figure}

We compare the theoretical charged particle pseudorapidity distribution (open circles)
in 0-6\% most central Au+Au collisions at $\sqrt{s_{\rm{NN}}}$=200 GeV with the
corresponding PHOBOS data (solid squares) \cite{phob2} in Fig.~1~(a). One sees here
that the PHOBOS data are well reproduced. In Fig.~1 (b), we compare the charged
particle pseudorapidity distributions in 0-5\% (open circles) and 5\% (open triangles)
most central Au+Au collisions with the 0-10\% one (open squares). We see in Fig.~1 (b)
that the pseudorapidity distribution in 5\% most central collision is quite close to
the 0-10\% one, because the 5\% centrality is nearly equal to the average centrality
of 0-10\% centrality bin.

\begin{figure}[htbp]
\includegraphics[height=2.5in,angle=0]{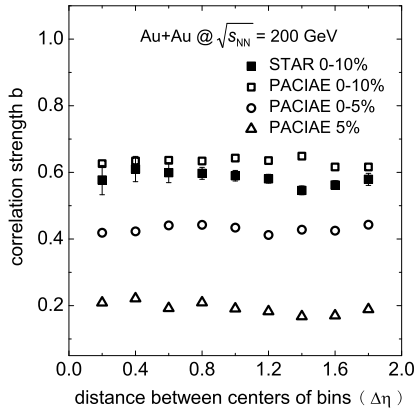}
\caption{Charged particle forward-backward multiplicity correlation
strength $b$ in 0-10, 0-5, and 5\% most central Au+Au collisions at
$\sqrt{s_{\rm{NN}}}$=200 GeV. The experimental data are taken from
\cite{star3}.} \label{fig2}
\end{figure}

In Fig.~2 we compare the calculated real (total) correlation strength $b$ (open
squares) as a function of $\Delta \eta$ in 0-10\% most central Au+Au collisions at
$\sqrt{s_{\rm{NN}}}$=200 GeV with the corresponding STAR data (solid squares)
\cite{star3}. The STAR data feature of correlation strength $b$ is approximately flat
across a wide range in $\Delta \eta$ are well reproduced. For comparison we also give the
real (total) correlation strength in 0-5 and 5\% most central collisions by open circles
and triangles, respectively. One sees here that the real (total) correlation strength
decreases with decreasing centrality bin size monotonously, because the charged
particle multiplicity fluctuation decreases from 0-10 to 0-5 and to 5\% monotonously,
as one will see in Fig.~3. This first result of the correlation strength increases with
increasing centrality bin size monotonously given in the transport model remains to be
proved experimentally.

\begin{figure}[htbp]
\includegraphics[width=7cm,angle=0]{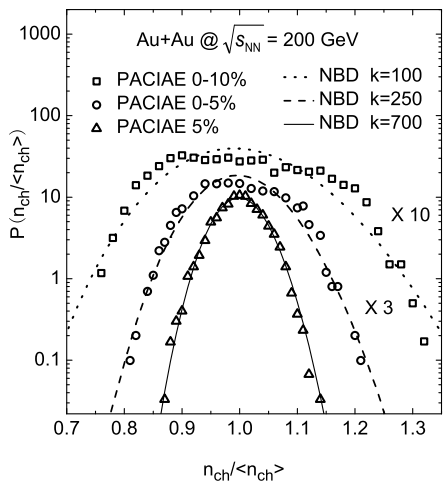}
\caption{Charged particle multiplicity distributions in 0-10 (open
squares), 0-5 (open circles), and 5\% (open triangles) most central
Au+Au collisions at $\sqrt {s_{\rm{NN}}}$=200 GeV. The dotted,
dashed, and solid lines are the corresponding NBD fits,
respectively.} \label{fig3}
\end{figure}

The calculated charged particle multiplicity distributions in 0-10, 0-5, and 5\% most
central Au+Au collisions at $\sqrt{s_{\rm{NN}}}$=200 GeV are given in Fig.~3,
respectively, by the open squares, circles and triangles. The corresponding NBD fits
are shown by dotted, dashed, and solid lines, respectively. One sees in Fig.~3 that
the charged particle multiplicity fluctuation is increased and the NBD fit is worsened
with increasing centrality bin size monotonously.

\begin{figure}[tb]
\includegraphics[width=7cm,angle=0]{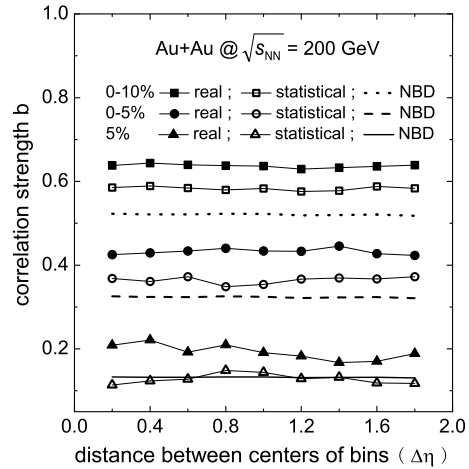}
\caption{The calculated charged particle total (real), statistical,
and NBD correlation strengths in 0-10, 0-5 and, 5\% most central
Au+Au collisions at $\sqrt{s_{\rm{NN}}}$=200 GeV.} \label{fig4}
\end{figure}

In Fig.~4 we compare the calculated charged particle real (solid symbols), statistical
(open symbols), and NBD (lines) correlation strengths as a function of $\Delta \eta$
in 0-10, 0-5, and 5\% most central Au+Au collisions at $\sqrt{s_{\rm{NN}}}$=200 GeV.
The solid squares, open squares, and dotted line are for 0-10\% most central
collisions, solid circles, open circles, and dashed line for 0-5\%, and solid
triangles, open triangles, and solid line for 5\%, respectively. We see in Fig.~4 that
the behavior of correlation strength increases with increasing centrality bin size
monotonously is not only existed in the real correlation strength but also in the
statistical and NBD ones.

If the discrepancy between real (total) and statistical correlation strengths is
identified as the dynamical correlation strength, one then sees in Fig.~4 that the
dynamical correlation strength may just be a few percent of the total (real)
correlation strength. The dynamical correlation strength in 0-10\% most central
collision is close to the one in 5\% most central collision globally speaking. That
is because the later centrality is nearly the average of the former one. The
dynamical correlation strength in 0-10\% most central collisions is globally less
than 0-5\% most central collision. That is because the interactions (represented by
the collision number for instance) in the former collisions is weaker than the later
one. We also see in Fig.~4 that the statistical correlation strength is nearly the
same as the NBD one in the 5\% most central collision, that is consistent with the
results in $p+p$ collisions at the same energy \cite{yan}. However the discrepancy
between statistical and NBD correlation strengths seems to be increased with
increasing centrality bin size monotonously. That is mainly because the NBD
fitting to the charged particle multiplicity distribution becomes worse with
increasing centrality bin size monotonously.

\section {CONCLUSION}
In summary, we have used a parton and hadron cascade model, PACIAE, to study the
centrality bin size dependence of charged particle forward-backward multiplicity
correlation strength in 5, 0-5, and 0-10\% most central Au+Au collisions at $\sqrt{s_
{\rm{NN}}}$=200 GeV. The real (total), statistical, and NBD correlation strengths are
calculated by real events, mixed events, and NBD method, respectively. The corresponding
STAR data feature of the correlation strength $b$ is approximately flat across a wide
range in $\Delta \eta$ in most central Au+Au collisions is well reproduced. It is turned
out that the correlation strength increases with increasing centrality bin size
monotonously. This first result, given in the transport model, remains to be proved
experimentally. If the discrepancy between real (total) and statistical correlation
strengths is identified as dynamical one \cite{yan}, then the dynamical correlation may be
just a few percent of the total (real) correlation. As a next step, we will investigate
the relation between correlation strength $b$ and the centrality bin size in the
mid-central and peripheral collisions, and the STAR data feature of $b$ approaches an
exponential function of $\Delta \eta$ at the peripheral collisions.

\begin{center} {ACKNOWLEDGMENT} \end{center}
Finally, the financial support from NSFC (10635020, 10605040, and 10705012) in China is
acknowledged

\end{document}